\documentclass[epj]{svjour}
\usepackage{graphics}
\usepackage{graphicx}
\usepackage{epstopdf}
\begin{document}
\title{Magnetism and unusual Cu valency in quadruple perovskites} 
\subtitle{}
\author{Paola Alippi\inst{1} and Vincenzo Fiorentini\inst{2}}\institute{CNR-ISM, Istituto di Struttura della Materia, Consiglio Nazionale delle Ricerche, Via Salaria km 29.5 CP 10, 00016 Monterotondo Stazione, Italy \and Dipartimento di Fisica, Universit\`a di Cagliari and CNR-IOM, UOS Cagliari, Cittadella Universitaria, Monserrato, 09042 Cagliari, Italy} \date{Date to be inserted}
\abstract{We study a selection of  Cu-containing magnetic quadruple perovskites (CaCu$_{3}$Ti$_{4}$O$_{12}$
LaCu$_{3}$Fe$_{4}$O$_{12}$, and
YCu$_{3}$Co$_{4}$O$_{12}$) by ab initio calculations, and show that Cu is in an effective divalent Cu(II)-like state or a trivalent Cu(III) state depending on the choice of octahedral cation. Based on the electronic structure, we also discuss the role of Mott and Zhang-Rice physics in this materials class.  
}
\authorrunning{P. Alippi and V. Fiorentini}
\titlerunning{Magnetism and unusual Cu valency in quadruple perovskites }
\maketitle
\noindent
\section{Introduction and methods}

The quadruple perovskites ACu$_{3}$X$_{4}$O$_{12}$ have attracted mu\-ch interest recently with several puzzling  behaviors such as anomalous dielectric response  \cite{ccto},  isostructural metal-in\-su\-lator phase transitions  \cite{lacufeo}, the cation selectivity of metallic or insulating character  \cite{cacucoo}, and, not least, the unusual valence of Cu  depending on the A and X cations. In this paper we analyze the latter two aspects drawing on ab initio density-functional calculations for a selection of different materials in this class, specifically for the A,X pairs (Ca, Ti), (La, Fe), (Y, Co), i.e. for 
CaCu$_{3}$Ti$_{4}$O$_{12}$
LaCu$_{3}$Fe$_{4}$O$_{12}$, and
YCu$_{3}$Co$_{4}$O$_{12}$ (labeled CCTO, LCFO, and YCCO henceforth).

 The presence of active 3$d$ shells --especially that of Cu-- suggests a possible role of electron correlation in these materials. The use  of a beyond-(semi)local density-functional approach may be advisable or outright necessary. The front runners in this field today are hybrid \cite{hse} or self-inter\-action corrected \cite{vpsic} functionals, which both correct the dominant error of local functionals in dealing with localized states, namely, self-interaction. For computational simplicity, here we use GGA+U (generalized gradient approximation plus ``Hubbard U''), which in this context  can be viewed as an approximate parametrized self-interaction correction for a specific orbital shell. While not especially satisfactory compared to more refined methods, GGA+U is quite sufficient in this context. Along the same line of correlation-related properties, we also preliminarily discuss  the role of ''Zhang-Rice'' physics \cite{cacucoo}  in these materials.

We use the PAW method \cite{paw} as implemented in the VASP code  \cite{vasp}, with the Dudarev GGA+U functional  \cite{dudarev}. We use  4$\times$4$\times$4 k-point meshes and cutoff energy 400 eV.
The U--J corrections applied to the 3$d$ states  are literature values   for Fe (5.4 eV, used for LaFeO$_{3}$ \cite{lafeo}) and Co (6.9 eV, used for LaCoO$_{3}$ \cite{lacoo}). For Cu in CCTO we use U--J=4 eV, similar to that used in Ref.\cite{noi}.  

\section{Results and discussion}
\subsection{Background}
The properties of this class of perovskites vary according to the choice of the A and X cations. 
The cubic ``A-type'' perovskite sublattice is occupied by Cu ions binding four oxygens into mutually orthogonal plaquettes  and by A cations sitting on an ordered BCC sub-sublattice. The X cations occupy the octahedrally-coordinated perovskite ``B'' sites. 
The metallic or insulating behavior can be rationalized largely based on sum rules for the nominal ionic valencies and the number of active electrons.  In an ionic picture based on oxidation numbers,  one expects an ionic insulator if
$$n_{\rm A} + 3n_{\rm Cu} + 4n_{\rm X} -12 n_{\rm O}=0,$$
$n$ being the valency of each of the species involved. 
  O is assumed to be a  nominally divalent anion, i.e. $n_{\rm O}$=--2. Cu can have valence $n_{\rm Cu}$=1, 2, or 3. If A=Ca (a divalent element), $n_{A}$=2, and hence
  \begin{equation}
3n_{\rm Cu} + 4n_{\rm X}=22,
\label{eq1}
\end{equation}
which is  satisfied by $n_{\rm Cu}$=2 and $n_{\rm X}$=4.  On the other hand, $n_{\rm A}$=3 if A=La or Y (both trivalent elements),  and
\begin{equation}
3n_{\rm Cu} + 4n_{\rm X}=21,
\label{eq2}
\end{equation}
which is  satisfied by $n_{\rm Cu}$=$n_{\rm X}$=3. 

A supplementary point of view is provided by the total valence electron count. If the latter is even, a band insulator can be realized. As shown below, this is indeed the case of LCFO and YCCO, with magnetism playing a secondary role. If the count is odd, additional symmetry breaking will have to occur, such as magnetic moment formation, and the resulting  insulator may be categorized as a Mott insulator. In the present context, this peculiarity is caused by the nine electrons of the  Cu(II) cation, and indeed occurs in CCTO, where magnetic polarization of Cu is essential to open a gap, as discussed below.  Viceversa, for instance, CaFe$_{3}$Ti$_{4}$O$_{12}$, where Fe plays the ``divalent'' role of Cu, has even electron count and is technically a band insulator (although in fact magnetic) \cite{cafetio}.

Assuming now that X is either a tetravalent cation such as Ti, or a  trivalent ion such as Fe and Co, we can easily set up examples of instances of Eqs.\ref{eq1} and \ref{eq2}. For Eq.\ref{eq1},  assuming Cu behaves as divalent, we obtain CaCu$_{3}$Ti$_{4}$O$_{12}$, which is indeed insulating \cite{noi,vanderbilt,noi2}, as discussed below.  Cu is 2+ and $d^{9}$, which as mentioned requires magnetic symmetry breaking. Examples of Eq.\ref{eq2} are LaCu$_{3}$Fe$_{4}$O$_{12}$ and YCu$_{3}$Co$_{4}$O$_{12}$; since in this context Fe and Co are  trivalent ions, assuming also a   trivalent Cu(III), both LCFO and YCCO should be, and indeed are, insulating, both experimentally \cite{lacufeo,cacucoo} and theoretically (as shown below). 
While Cu(III) is known to occur infrequently, our results support  its  occurrence in the materials at issue here, although with inevitable  hybridization with O and the X cation.

\begin{figure}[h]
\center\includegraphics[width=7 cm]{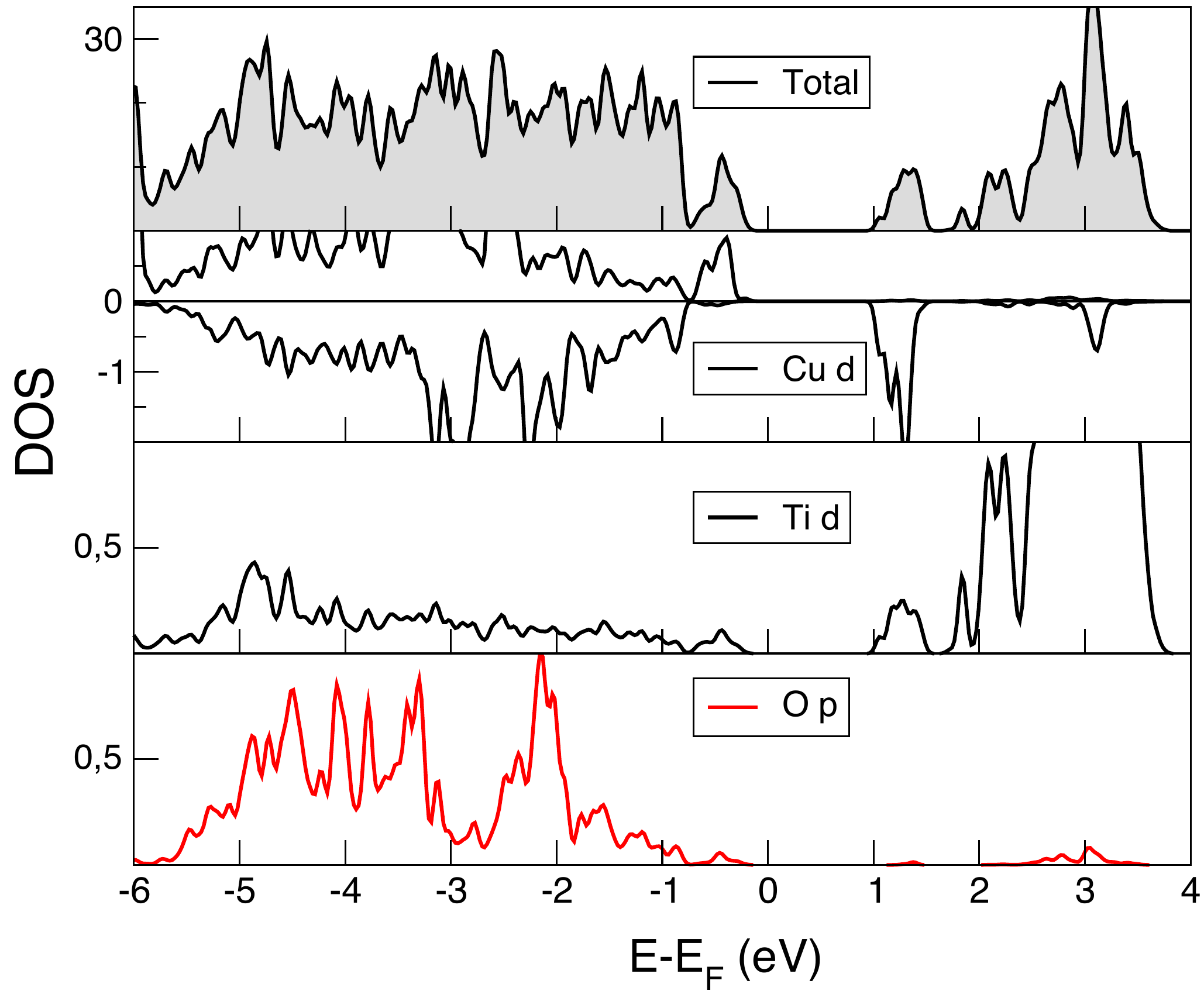}
\caption{Density of states of CCTO from GGA+U \protect\cite{futuro}.}
\label{fig2}
\end{figure}

\begin{figure}[h]
\center\includegraphics[width=7 cm]{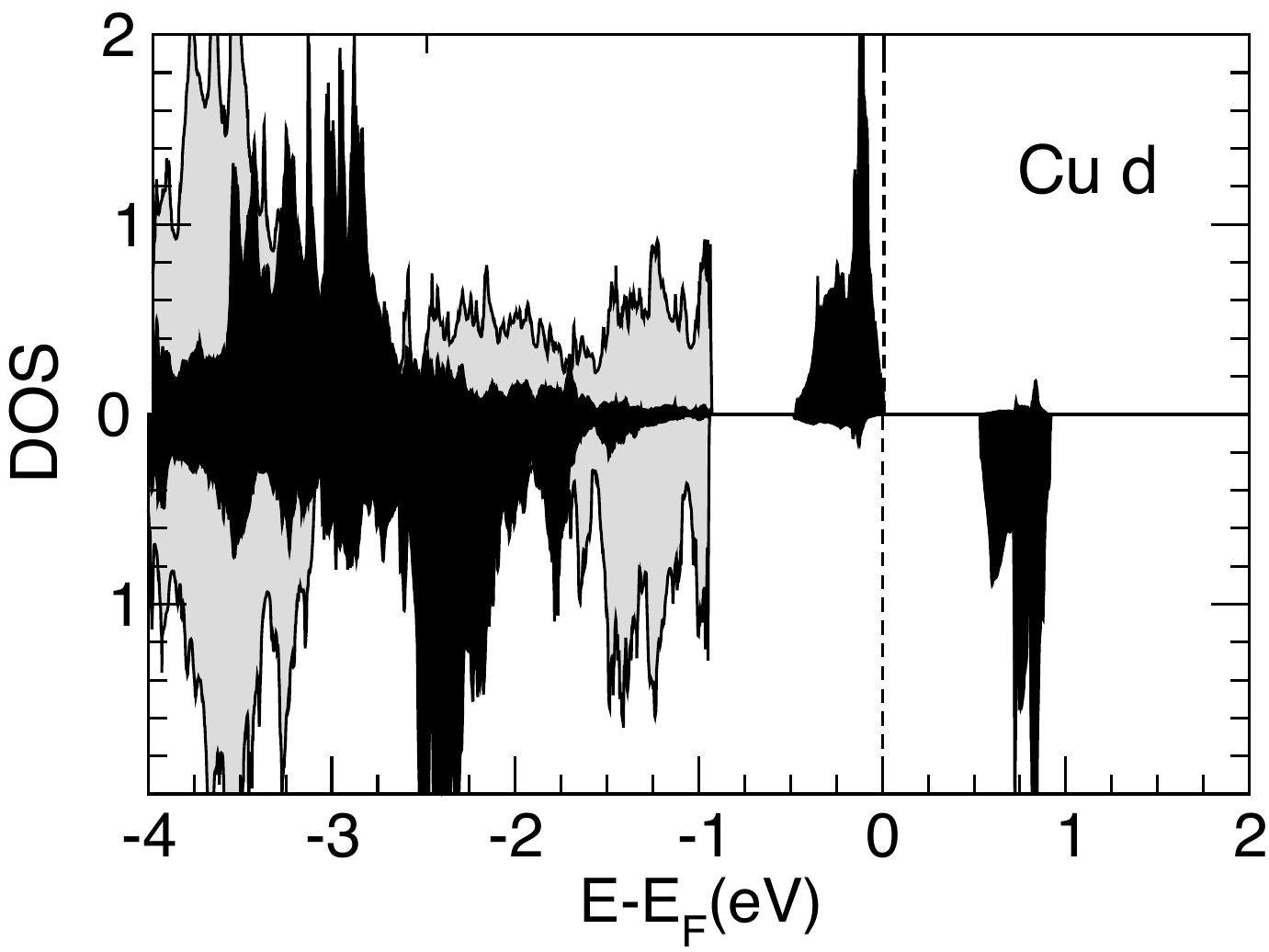}
\caption{Cu 3$d$ (black: $t_{2g}$; gray: $e_{g}$) density of states of CCTO obtained from self-interaction-corrected LDA \protect\cite{futuro}}
\label{fign}
\end{figure}
One variant of this idea is that a nominal Cu(III) state may correspond to  $d^{9}$+$\underline{L}$ (one native Cu hole and one hole on ligand oxygens) rather than to a proper $d^{8}$ state.   Such a coupled two-hole object is usually dubbed a ``Zhang-Rice'' state \cite{zr}. We discuss the concept below, pointing out that CCTO and LCFO are not Zhang-Rice-like themselves, but may be extremal points of a Zhang-Rice-like doping series. 

For the present  purposes it is sufficient to adopt the experimental  lattice constants (7.38 \AA\, for CCTO, 7.43 \AA\, for LCFO,   7.12 \AA\, for YCCO). The calculated values are, in fact, within 1\% of experiment or closer.
Internal coordinates (quite similar in all cases, given that Cu-O plaquette formation largely determines the rotations) are optimized. 

\subsection{Electronic structure}
The density of states (DOS) of CCTO is shown in Fig.\ref{fig2}. The main feature   is the narrow top valence state  of  majority-spin Cu-like character. This stems from a singlet a$_{g}$ Cu state in the plane of the CuO$_{4}$ plaquette,  with $t_{2g}$ orbital character   for our  choice of the local cartesian axes. The bottom conduction state is the minority-spin state with the same character. The fundamental gap is 1 eV, indirect and dipole-forbidden.  The higher transitions are O $p$--Ti $d$ at over 2.5 eV.  This is roughly  consistent  with reflectivity \cite{jacs} and optical conductivity \cite{lunk}.  As in most cuprates, Cu magnetic moments (calculated  inside spheres of appropriate radii) are about 0.6 $\mu_{\rm B}$, signaling important hybridization with O.

Let it be mentioned that  the electronic properties of CCTO are somewhat more complicated than briefly outlined above. A detailed comparison of hybrid-functionals, GW, GGA+U, and self-inte\-rac\-tion correction results will be presented elsewhere \cite{futuro}.  For one thing, the U value of 4 eV,  appreciably smaller than the fairly usual 8-9 eV used for cuprates \cite{ybcochain} such as YBCO, still appears to push the empty and filled Cu $d$ states too far apart: for comparison we show in Fig.\ref{fign} the Cu $d$ DOS computed \cite{futuro} by variational self-inte\-rac\-tion corrected LDA \cite{vpsic}. Clearly the lowest gap is smaller and the Cu states are more isolated from the main covalency-driven valence (and conduction, not shown) bands.  
In any event, the point in question here (Cu effective valency)  is not influenced significantly by these details.

\begin{figure}[h]
\center\includegraphics[width=8 cm]{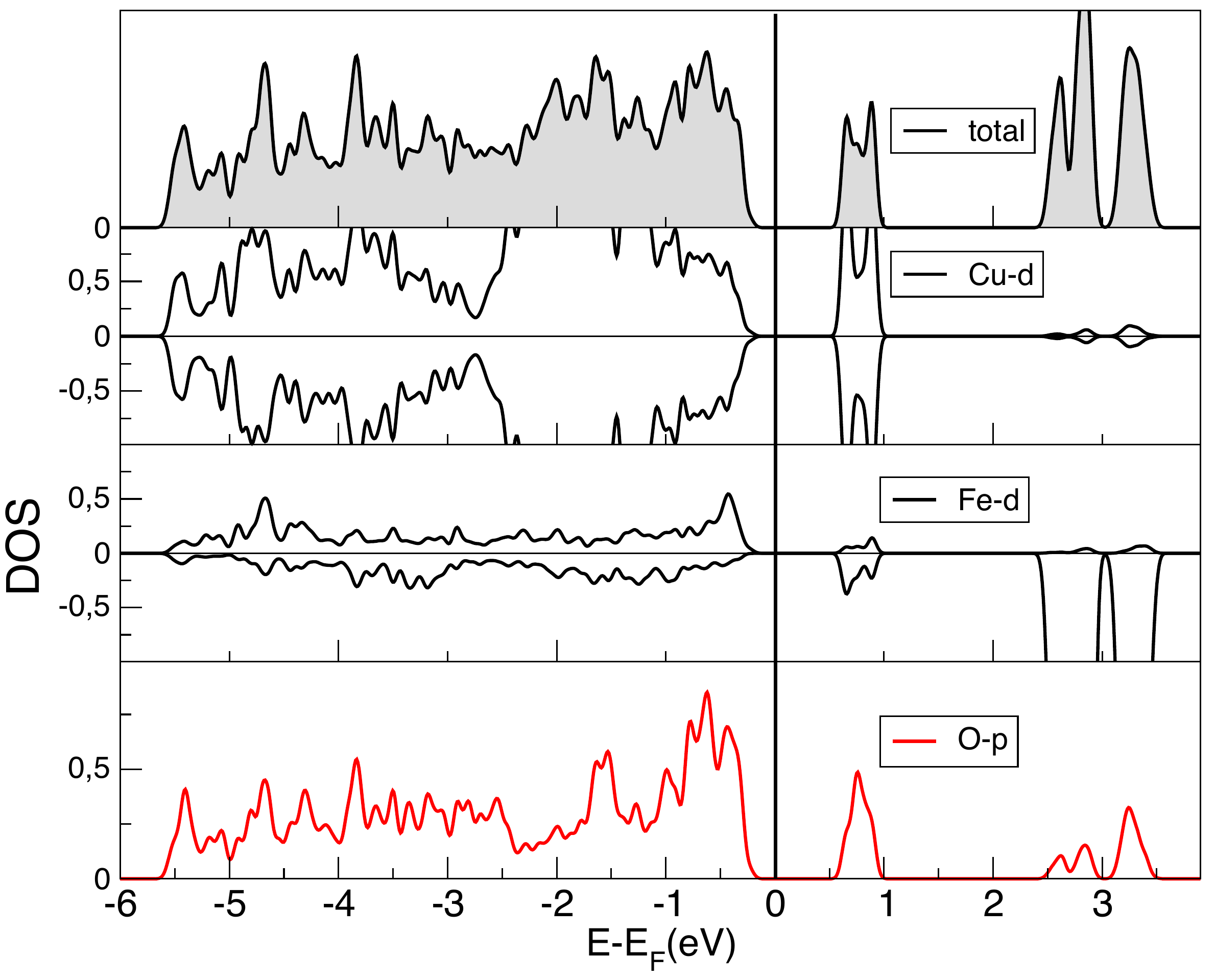}
\caption{Total and site-projected density of states for LCFO.}
\label{fig3}
\end{figure}

The DOS of LCFO is displayed in Fig.\ref{fig3}. The ground state order is antiferromagnetic G-type on the Fe lattice, with moments of 4.2 $\mu_{\rm B}$, and zero moments on Cu. Moments on O are all zero as well.  This result was validated by hybrid-functional calculations (not presented here). A fairly extensive search (by hybrid functionals and GGA+U)  for a ground state  combining moments on both Fe and Cu was unsuccessful, while uncovering a number of  metallic  magnetic excited configurations. Metallicity is in fact not unexpected in such cases, since Cu-Fe interactions suffer considerable frustration in this geometry. Clearly, our ground state is  consistent with trivalent Fe(III) in a nominal $d^{5}$ t$^{3}_{\uparrow}$e$^{2}_{\uparrow}$ high-spin state, and with non-magnetic Cu  in a  $d^{8}$  state. The electronic structure confirms this conclusion: the gap, again a little over  1 eV, now opens between a mostly O-p top valence band and the empty Cu a$_{g}$ orbital singlet (and degenerate spin doublet). This state was spin-split, polarized, and half-occupied in CCTO.  As in the latter, higher  O $p$--Fe minority-$d$ transitions would start only above 2.5 eV.  Interestingly, confirming the band insulator nature of LCFO, even standard GGA finds a small gap in LCFO.

\begin{figure}[ht]
\center\includegraphics[width=8 cm]{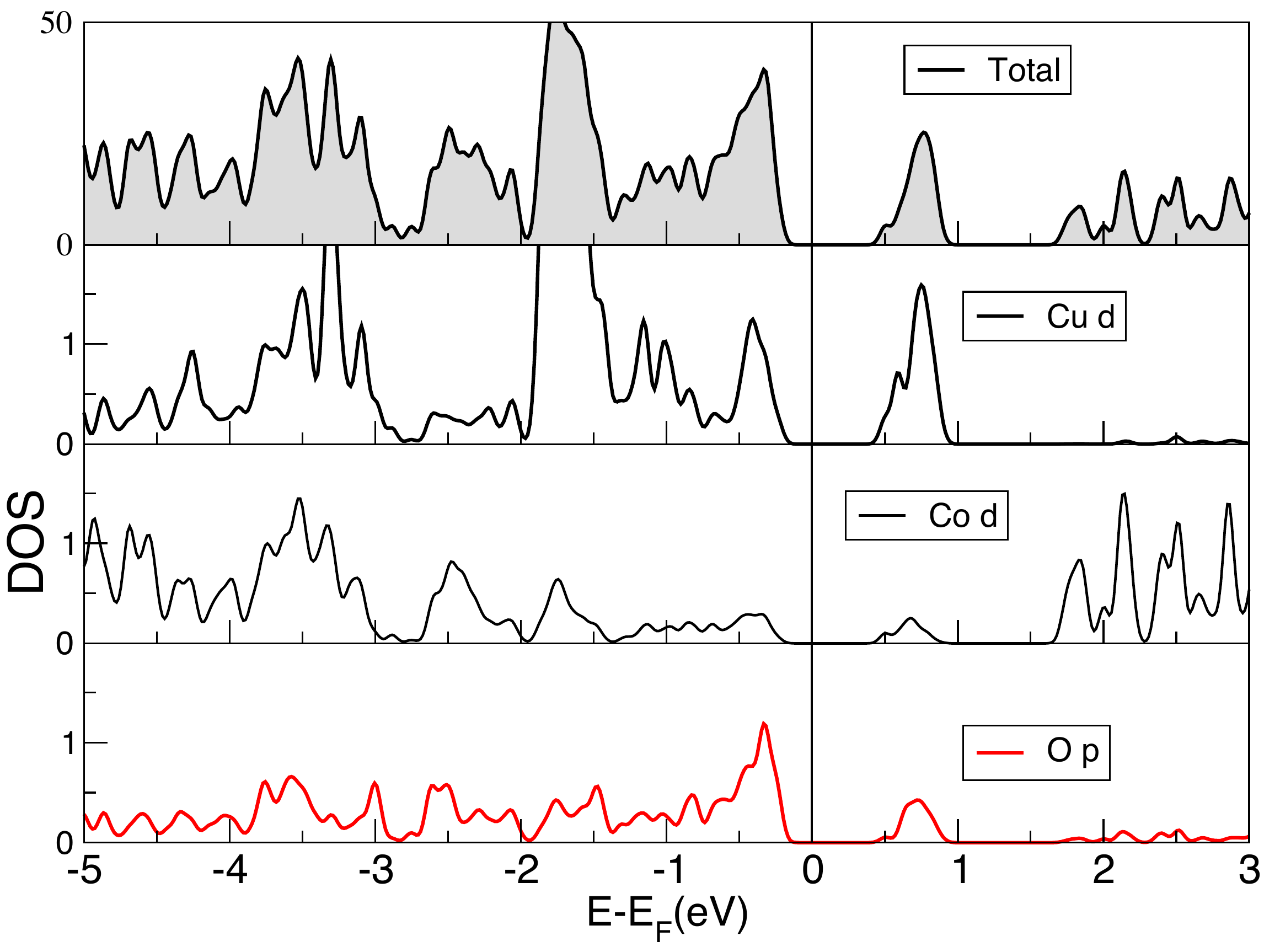}
\caption{Total and site-projected density of states for YCCO.}
\label{fig4}
\end{figure}

For  YCCO, we sampled a number of  magnetic configurations. Only one of these was found to be insulating (as found in experiment \cite{cacucoo}): the non-magnetic state, with zero moments on both Co and Cu. In this phase as well as in metallic ones with polarized Co, we also invariably found Cu to end up in a non-magnetic state irrespective of the starting point. We  thus conclude that both Co and Cu are in a low (in fact zero) spin state, respectively  Co:$d^{6}$ t$^{3}_{\uparrow}$t$^{3}_{\downarrow}$ and Cu $d^{8}$ as in LCFO. This non-magnetic state has been found \cite{lacoo} to compete energetically with an intermediate spin state in LaCoO$_{3}$ in LDA+U calculations similar to ours. Indeed, we find an analogous metallic state to be slightly lower in energy than the insulating paramagnet. Below, however, we focus on the latter, in view of its insulating character and of the uncertainties in the energetics due to the use of multiple U's.

The gap is about 0.6 eV, between mostly O $p$ (although Cu- and Co-admixed)  states and a mainly Cu-like empty orbital singlet (degenerate spin doublet). Higher transitions of O $p$--Co $d$ character start, once more, at about 1.8 eV.  %
An interesting point is that, as mentioned, the electron count is even in YCCO, which  could then be a gap insulator even with no magnetic order. In GGA, however, non-magnetic YCCO is a metal. This must be attributed to the strong spurious on-site repulsion acting on Co $d$ states due to self-interaction. 
%
%
GGA+U enhances the t-e orbital polarization already present in GGA, and opens a gap without magnetism involved (we find that hybrid functionals produce the same qualitative effect). Thus, correlations beyond semi-local functionals as provided by GGA+U and hybrids (essentially, self-interaction removal) are needed to get the  on-site interaction energy right in YCCO, which may therefore  be termed a correlated band insulator (in the common, albeit questionable connotation of ``needing correlations beyond semilocal DFT'').

\subsection{Cu valence and moments}
The vanishing Cu moment and the Cu-dominated DOS of the lowest conduction (empty)  states are probably the best validation of  an effectively Cu $d^{8}$, i.e. Cu(III), state in LCFO and YCCO, especially in comparison with the lonely, singly-occupied majority Cu(II) state in the lower part of the CCTO gap. 
%
%
%
A direct assessment of the charge residing on individual constituent atoms would lend  further support to this claim. However such assessment is  impossible by definition in multi-atom quantum systems. (Many ingenious methods have been used to approximate it \cite{bader,lanimno}). 
In partially covalent crystals, massive hybridization further blurs  the attribution of charge population to individual atoms and species. Nevertheless, to  provide a semi-quantitative comparison of the\, ``active'' charge sitting on Cu in LCFO and CCTO, we choose to integrate the Cu-projected DOS of the uppermost valence state and the lowermost conduction state  in CCTO, and the Cu-project\-ed DOS of the lowest conduction state in LCFO. We reckon in fact that attempting to account for changes in the whole atomic charge (i.e. integrating the  DOS over all energies) would swamp the detailed features we are looking for, essentially due to a natural charge back-flow upon population changes \cite{zunger} typical --but not exclusive-- of defects \cite{fegan}.     

In the CCTO case, the integral of each  of the peaks in  the Cu-projected DOS   would be unity for a perfect Cu(II) state; in LCFO and YCCO, the integral of the Cu-projected peak  would be also unity for an ideal Cu(III). In CCTO  we find an integral of 0.65 for the filled peak, in obvious correspondence with the Cu magnetic moment (Cu moments of 0.5-0.7 are quite usual in copper oxides due to the large hybridization with O).  The empty peak integrates to about 0.8. For the LCFO and YCCO empty Cu-like peak, about the same applies,  with an integral of about 0.5 (per spin channel). Most importantly,  the magnetizations on Cu and neighboring O in LCFO and YCCO are both zero as expected from a Cu(III) state, and not opposite and  compensating as would be the case for a $d^{9}$+$\underline{L}$ state.  We consider  this  to be  rather convincing evidence in favor of our suggestion  about the Cu(II) CCTO state and the Cu(III) LCFO / YCCO state.

We now propose a chemical rationale or the preference for Cu oxidation to Cu(III) in association with Fe(III) rather than to Cu(II) associated to Fe(IV) using   atomic oxidation potentials. We estimate the energetics  of oxidation combining the oxidation potentials of all species involved, and  decide accordingly which valency combination, Cu(III)/Fe(III) or Cu(II)/Fe(IV), is favored. Using standard values, or estimates from Frost diagrams \cite{atkins}, we find that, all else being equal, oxidation to Fe(II) is less costly than to Fe(IV) by 0.35 eV/Fe or 1.2 eV/formula unit in this stoichiometry. The tetravalent state of Co  is even more unfavorable (over 2 eV/Co). To compare two situations satisfying the ionic sum rule, in the estimates we used the La potential for the trivalent pair and the Ca potential for the divalent-tetravalent; this is however quite  immaterial as the Ca and La potential are almost the same. Further, we assumed that the energetics of ionic bonding and covalency effects do not change significantly replacing Fe(III)/Cu(III) with Fe(IV)/Cu(II). In  reality, in the latter case, we would probably have a Cu-$d$ Mott gap as in CCTO, further disfavoring the Fe(IV)/Cu(II) configuration.
   
Similar arguments applies to CaCu$_{3}$Co$_{4}$O$_{12}$ and Ca\-Cu$_{3}$\-Ru$_{4}$O$_{12}$, which fail to satisfy the ionic sum rule assuming trivalent Ru and Co.  Both systems may be insulating via a Mott mechanism similar to CCTO if  Co and Ru were tetravalent; instead they end up being metallic \cite{cacucoo,cacuruo},   which indicates that the fourfold oxidation of both Ru and Co is too costly energetically, and the system prefers to metallize, irrespective of Cu valence. (Also, a hypothetical sum-rule-satisfying mixture of Cu 2+ and 3+ would destroy the CCTO-like gap between Cu states.) CaCu$_{3}$Co$_{4}$O$_{12}$  turns out \cite{cacucoo} to be metallic and weakly ferromagnetic with moments of 0.6 $\mu_{\rm B}$ on  Cu and 0.87 $\mu_{\rm B}$   Co, which has thus a nominally intermediate valence 3+$\delta$ and low spin t$^{3}_{\downarrow}$t$^{3}_{\uparrow}$e$^{\delta}_{\uparrow}$ configuration, with $\delta$$\sim$0.2.

 \subsection{Zhang-Rice behavior ?}
  
 We conclude this paper discussing the role of  ''Zhang-Rice'' physics envisaged in recent experimental work \cite{cacucoo}, setting also the stage for future work. 
In this context, ''Zhang-Rice'' physics has to do with whether Cu(III) is  $d^{8}$  (two holes in the Cu $d$ shell) or rather $d^{9}$+$\underline{L}$, the hole $\underline{L}$ being on  ligand O's.  The namesake is the  Zhang-Rice  singlet polaron in cuprates \cite{zr}, a composite of the native hole of $d^{9}$ Cu and a dopant hole on ligand oxygens, which are oppositely spin-polarized and produce a locally unpolarized object.
Zhang-Rice-like behavior  in the form of localization phenomena at or near a Cu site may be expected when  an isolated, narrow Cu-like band is hole-doped between the nominal Cu(II) and Cu(III) states. To understand this in terms of band structure, a useful reference is an earlier study of Ca-doped Y cuprate \cite{caycuo}. At zero Ca content,  it  is insulating with a narrow upper valence band of strong Cu-like character, indeed commonly called  Zhang-Rice band in the literature. A quite similar band, we have seen above, is present  in  CCTO. As  doping is increased, the electronic structures exhibits typical strong-correlation phenomena  such as spectral weight rearrangement, spontaneous localization, and multiple metal-insu\-la\-tor transitions. At full Ca-doping (one hole per Cu), the Zhang-Rice band is completely empty, and a small gap reappears (this is usually named a Kondo insulator). The latter situation is analogous to LCFO, where the majority Cu-like top valence band of CCTO is  empty and sits low in the O $p$-Fe $d$ gap. (For  the``Mott'' physics often invoked in this context, see \cite{mott}.)

First off,  our analysis suggests that  neither CCTO nor LCFO  properly belong in the Zhang-Rice variety. The full ``hole'' that Cu acquires going from CCTO to LCFO (i.e. the Cu $d$ band filled in CCTO and  empty in LCFO) is still largely localized on Cu, and what little sits on ligands is not spin-polarized. On the other hand,  CCTO and LCFO may be seen as the  end  points of the hole doping range of the flat Cu-like top-valence Zhang-Rice band of CCTO. This suggests --although it does not {\it imply}-- the possible appearance of Zhang-Rice-like polarons or other strong-correlation phenomena at intermediate doping. (Dynamical effects such as multiplets and satellites may be revealed even in the end-point materials by optical probes, effectively ``instantaneously doping'' the material.)  

The nominal  ``one-hole-per-Cu'' doping in LCFO compared to CCTO  is caused by Fe's  being 3+, hence retaining 4 more electrons than the Ti's in CCTO, and La giving one electron more than Ca -- hence overall one electron missing from each of the three Cu's.  Computationally, instead, one can dope CCTO in all the range from 0 to 1 hole/Cu by diminishing the number of electrons by 0 (CCTO) through 6 (LCFO) in the magnetic 40-atom (2 formula units) cell of CCTO, and assuming a compensating background. Work in this direction will be reported elsewhere. Experimentally, one could try  mixing in a trivalent X cation and/or substitute Na for Ca in CCTO. The former could attain any doping level, at least in principle, while the latter would be limited to a maximum  doping of 1/3 hole per Cu at full Na substitution for Ca, i.e. in NaCu$_{3}$Ti$_{4}$O$_{12}$.

\section{Summary}

In summary, we have discussed the electronic and magnetic properties of the three quadruple Cu-containing perovskites CCTO, LCFO, and YCCO. CCTO is a an Cu-site AF-G Mott insulator; LCFO is a Fe-site AF-G band insulator; YCCO is a non-magnetic correlated band insulator. Fe is in Fe(III) high-spin state; Co is in a Co(III) low-spin state; Cu is Cu(II) and magnetic in CCTO, and Cu(III) and non-magnetic in LCFO and YCCO. This is borne out both by  magnetic-state and charge-population analyses. We  discussed the (unimportant) role of Zhang-Rice physics in CCTO and LCFO, suggesting that CCTO and LCFO are end points of a potential Zhang-Rice doping series, and proposed  routes to produce it.

\section*{Acknowledgments} Work supported in part by projects EU FP7 {\it ATHENA} and IIT Seed {\it NEWDFESCM}, and by a Fondazione Banco di Sardegna 2011 grant. Computing resources provided by CASPUR Rome under competitive grants and by Cybersar Cagliari.


\begin{thebibliography}{99}
\bibitem{mott}
''Mott'' physics in this context refers to the system being insulating despite an odd electron count, predominantly due to on-site e-e repulsion. Most Mott insulators are magnetically ordered --generally antiferro-- at low temperature. This is the case of CCTO described above. At high temperature moments disorder, resulting in a paramagnet which is well  described by band theory methods, provided a beyond-local DFT approach is used (in conjunction with finite temperature multiple scattering \cite{luders}). To describe the Mott state,  ab initio methods must capture simultaneously the quite different energy scales set by direct electron-electron interaction (Hubbard U$\sim$1-10 eV), variants of exchange (0.1-1 eV, involving Hund's rule, superexchange, double and kinetic exchange, etc.), and detailed geometric structure (0.01-0.1 eV). This is currently only possible, with ample qualifications, via approaches such as GW, hybrid functionals, and self-inte\-rac\-tion corrections, which share as a common element  the removal of the self in\-te\-rac\-tion present in  the (semi)local approximations to the density functional. Despite their low energy scale,   structural details play a key role: oxides of degenerate open-shell-atoms such as Cu(II) $d^{9}$ or Ti(III) $d^{1}$ invariably resort to octahedra rotations and Jahn-Teller distortions, coupled to magnetism, to remove electronic degeneracies and produce insulating states \cite{vpsic,cuo}.
\end{thebibliography}
\end{document}